\documentclass[12pt]{article}
\newcommand{\sect}[1]{\setcounter{equation}{0}\section{#1}}

\evensidemargin=\oddsidemargin
\newcommand{\be}{\begin{equation}}
\newcommand{\ee}{\end{equation}}
\newcommand{\bea}{\begin{eqnarray}}
\newcommand{\eea}{\end{eqnarray}}
\newcommand{\p}{\partial}
\newcommand{\al}{\alpha}
\newcommand{\bt}{\beta}

\begin{document}
\renewcommand{\thefootnote}{\fnsymbol{footnote}}
\begin{titlepage}
\begin{flushright}
ROM2F/2004/19\\
hep-th/0407134\\
\end{flushright}
\vskip .7in
\begin{center}
{\Large \bf D-Brane Dynamics in Dp-Brane Background}
\vskip .7in
{\large Kamal L. Panigrahi}\footnote{e-mail: {\tt
Kamal.Panigrahi@roma2.infn.it}}
\footnote{\it INFN fellow}\\
\vskip .2in
{}{\it Dipartimento di Fisica, Universita' di Roma ``Tor Vergata"\\
INFN, Sezione di Roma ``Tor Vergata", Via della Ricerca Scientifica 1\\ 
00133 Roma, Italy} 
\vspace{.7in}
\begin{abstract}
\vskip .5in
\noindent
By using Dirac-Born-Infeld action we study the real time dynamics
of $D$-branes in the vicinity of a stack of $Dp$-branes where the role of 
the tachyon
of the open string models is played by the radial mode on the $D$-branes.
We examine the behaviour of the tachyon potential and study the hamiltonian
formulation and classical solutions of such systems. We also study the
homogeneous solutions of the classical equations of motion in these cases.
\end{abstract}
\end{center}
\vfill

\end{titlepage}
\setcounter{footnote}{0}
\renewcommand{\thefootnote}{\arabic{footnote}}
\sect{Introduction}

In the study of tachyon dynamics, in open string theories, construction
of various time dependent classical solutions has been an important
area of research in the recent past. These solutions represent the decay
of an unstable D-brane or a brane-antibrane pair as the tachyon rolls 
towards the minimum of the potential\cite{sen-RT}. The open string field 
theory on these systems have a natural time dependence and they are shown 
to be
spatially homogeneous classical solutions of arbitrary low energy density
\cite{sen-RT,sen-TM,sen-FT,sen-TE,sen-TT,GS,ST,minahan,hashimoto,sen-DD}.  
These solutions, in general, are constructed by perturbing the boundary 
conformal field theory that describes the D-brane by an exact marginal 
deformation. The real time tachyon dynamics shows that effective action
for the Dirac-Born-Infeld type \cite{sen-NBPS,garousi,BRWEP,Kluson}
captures many aspects of 
rolling tachyon solutions of the full string theory. The rolling tachyon 
solutions predicts that at late times tachyon condensation leads to a
tachyon matter state which has the equation of state similar to that of
a pressure-less fluid, and also there is no on-shell open string modes 
present. The open string picture of this process of time evolution 
leads to a new kind of open-closed string duality\cite{sen-OCD,sen-OCDM}.
In \cite{sen-TT}, it has been shown that in the theory of Born-Infeld 
coupled to gravity, 
at late time, the tachyon field $T$ appearing in the action could serve
as the definition of time for the canonical gravity problems. So the
speculation is that the tachyon field $T$ could also lead to the 
identification of an intrinsic time variable in the full string 
theory. A lot of work has also been dedicated in the analysis and 
the relevance of tachyon condensation in cosmological applications
\cite{gibb-cosmo,linde,fair-bairn,mukhoyama,padmanavan,burg,ghoshal,
sen-COSM,Sami}.
 
However in a recent paper by Kutasov \cite{kutasov} this
problem has been analyzed by using the Dirac-Born-Infeld (DBI) action
of a $Dp$-brane in the vicinity of $N$ number of coincident NS5-branes.
Though the $D$-branes
are stable but in the presence of NS5-branes they become unstable,
as they break all space time supersymmetries. 
So it is probably natural to think of this process as an example
of the tachyon condensation and study the closely related time 
evolution process of tachyon and the nature of the time dependent
solutions. These has been analyzed beautifully in 
\cite{kutasov} where the role of tachyon in this whole
process is played by the radial mode on the $Dp$-brane. 
Among the other things,
it was also argued and shown that the pressure goes to zero at late times
like $\exp(-\alpha t)$, where $\alpha$ depends on the number of five branes
and on the angular momentum of the $Dp$-brane. NS5-brane background is 
special in the following sense. First, there is no background RR flux
(hence the absence of WZ terms in the DBI action) and the harmonic 
function that describes the NS5-brane supergravity solution, i.e. 
$H = 1 + {N l^2_s\over R^2}$, makes the form of the tachyon potential  
grows exponentially as $R\rightarrow 0$ (where $R$ denotes the radial 
mode along the transverse directions of the $NS5$-branes). 
Second, the radial DBI action
interpolates smoothly between the standard gravitational interaction
between the $D$-brane and NS5-branes at large distances 
and a ``radion matter''
phase at a short distance between them \cite{kutasov}.  
This particular feature is absent in all the other $Dp$-branes.
Only in  the S-dual picture of the above, in the region $Ng_s \gg 1$,
one could examine these effects. In all other cases, as has been 
shown \cite{burgess1,burgess2} the
potential behaves power like and in some cases (e.g. a $D0$-brane 
moving in the vicinity of $D6$-branes, and when the probe brane and the 
background are of same dimension and parallel to each other etc.) 
one needs to include the WZ term in the full effective action. 
We, however, will not discuss those cases in this article.
Also when there are more that 
one probe brane then one needs to include the Myers type of 
effects \cite{myers}. But still analyzing carefully the classical
dynamics of branes in $Dp$-brane backgrounds, and studying the
time evolution process, one learns more about the nature of the 
time dependent solutions in string theory. 

In view of the recent developments in understanding better the classical
time dependent solutions, in this paper, we study the 
problem of the dynamics of $D$-brane near a stack of 
$Dp$-branes by mapping it to that of a tachyon effective action of open 
string models and study some of its consequences. 
The rest of the paper is organized as follows.
In section-2 we study the effective action, the effective potential
and the hamiltonian formulation for the system of a single probe brane in the
background of a stack of static and parallel branes. In section-3, we
study the homogeneous solutions of the equations of motion,         
and compute some interesting quantities like the stress
tensors. In section-4, we turn our attention to somewhat interesting
and closely related case of brane dynamics in the background of
$D5$-branes and examine the tachyon effective potential. Finally, in section-5 
we conclude with some discussion.

\sect{Effective action and Hamiltonian Formulation}
In this section, we analyze the motion of a $Dp'$-brane in the background 
generated by a stack of coincident and static $Dp$-branes, by using the
DBI action.\footnote{We consider the branes to be static, and ignore the
massive closed-string modes. Hence the low-energy expansion is valid
(neglecting $\alpha'$ corrections). These modes become relevant at $r<<l_s$.
But here we assume: $l^{7-p}_s << r^{7-p} << N g_s l^{7-p}_s$
\cite{kutasov}. The DBI analysis is perfectly valid in this 
domain and we can study the dynamics of branes in terms of 
effective action.} 
The metric, the dilaton $(\phi)$ and the R-R field $(C)$ 
for a system of $N$ coincident $Dp$-branes is given by:
\bea
g_{\al \bt} &=& H^{-{1\over 2}}_p ~\eta_{\al \bt}, 
\>\>\>\>  
g_{mn} = H^{1\over 2}_p ~\delta_{mn}, \>\> 
(\al, \bt = 0,..,p;\>\> m, n = p+1,..., 9), \cr
& \cr
e^{2\phi} &=& H^{{3-p}\over 2}_p, \>\> C_{0...p} = H^{-1}_p,\>\>\>\>  
H_{p} = 1 + {{N g_s l^{7-p}_s}\over r^{7-p}}.
\label{Dp-metric}
\eea
where $H_p$ is the harmonic function of $N$ $Dp$-branes satisfying the
Green function equation in the transverse space. In the subsequent analysis
we assume (1)~$p > p'$, (2)~there is no magnetic flux on the branes,
and (3)~there is a single probe brane at a time.
Since $p\ne p'$ and there is no
worldvolume gauge field on the probe brane, the R-R
fields don't affect the movements of the branes and they are only affected by
the gravity and by the dilaton.

The effective action on the world volume of $Dp'$-brane is governed by the
DBI action:
\begin{eqnarray}
S_{p'} = -T_{p'}
\int d^{p'+1} \xi e^{-(\phi -\phi_0)}\sqrt{-\det(G_{ab} + B_{ab})}.
\end{eqnarray}
Where $G_{ab}$ and $B_{ab}$ are the induced metric and the B-field,
respectively, on the world volume of the $Dp'$-brane:
\begin{eqnarray}
G_{a b} &=& {{\partial X^{\mu}}\over{\partial \xi^a}}
{{\partial X^{\nu}}\over{\partial \xi^b}}g_{\mu\nu}, \cr
& \cr
B_{a b} &=& {{\partial X^{\mu}}\over{\partial \xi^a}}
{{\partial X^{\nu}}\over{\partial \xi^b}}B_{\mu\nu},
\end{eqnarray}
where $\mu$ and $\nu$ runs over whole ten dimensional space time.
The worldvolume coordinates of $Dp'$-brane are leveled by $\xi^{a}$
$(a = 0,...,p')$,
and we set (by reparametrization invariance on the world-volume of the
$Dp'$-brane) $\xi^a = x^a$. The position of the D-brane in the
transverse direction gives rise to the scalars on its world-volume.
We restrict ourselves to the purely radial fluctuation along the
common $(9-p)$ transverse space $(R = \sqrt{X^m X_m (\xi_a)})$.
The induced metric on the worldvolume of the $Dp'$ brane is given by:
\bea
G_{ab} = H^{-{1\over 2}}_p~\eta_{ab} + H^{{1\over 2}}_p~\p_a R \p_b R.
\eea

So the DBI action of the $Dp'$-brane in the background generated by
$N$ $Dp$-branes is given by:
\bea
{\mathcal S}_{p'} = -\tau_{p'}\int d^{p' + 1}~x H^{{p-p'-4}\over 4}_p
\sqrt{1 + H_p \p_a R \p^a R}.
\label{brane}
\eea
The form of the above action looks somewhat similar to the DBI action 
of the tachyon field in the open string models, which is:
\bea
{\mathcal S}_{tach} = - \int d^{p' + 1}~x V(T) \sqrt{1 + \p_{a} T \p^a T}
\label{act-tach}
\eea
Comparing the above two actions, we can define a tachyon field $T$
by the following relation:
\bea
{dT \over dR} = \sqrt{H_p (R)} = \sqrt{1 + {{N g_s l^{7-p}_s}\over R^{7-p}}}.
\label{tachyon}
\eea
In terms of this field the ``tachyon potential'' in (\ref{act-tach}) 
is given by:  
\bea
V(T) = \tau_{p'} [H_p(R(T))]^{{p-p'-4}\over 4}
\eea
Before solving (\ref{tachyon}),
few remarks regarding the structure of the ``tachyon potential''
are in order now. One notices that for $p-p' = 4$, the
above potential is equal to the tension of the probe brane.
The would be tachyon mode is instead a massless scalar and the
static system is supersymmetric as it preserves 1/4 of the space time
supersymmetry (e.g. D1-D5 system). 
This supersymmetry is reflected in the absence of a
potential of interaction between the branes when they lie parallel
to each other and are static. When $p-p' < 4$ the potential is attractive
at short distances and therefore can form bound state
(e.g. non-threshold bound of D1-D3 branes). When $p-p'> 4$,
the effective potential is repulsive and the lowest 
open-string mode is massive at short distance. Many interesting  
facts regarding the behaviour of the effective potential in brane motions and
their classical orbits have been analyzed in great detail
in \cite{burgess1,burgess2}. 

The solutions of eqn. (\ref{tachyon}) is given by:
\bea
T(R) &=& -{2 R(1 + {{Ng_sl^{7-p}_s} \over{R^{7-p}}}) \over{5-p}} + {2(7-p)
\over{\sqrt{Ng_sl^{7-p}_s}}} {R^{{7-p +2}\over 2}
\over{{(7-p)}^2 - 4}} \times \cr
& \cr
&\times& {}_2F_1\left ( \frac{2+(7-p)}{2 \, (7-p)},
\frac{1}{2},1+\frac{2+(7-p)}{2 \, (7-p)}; -{R^{7-p}\over{N g_s l^{7-p}_s}}
\right ),
\eea
where ${}_2 F_1$ the hypergeometric function. Note that the above solution
is not valid for $p = 5$, where the form of the solutions and the
asymptotic behave differently \cite{kutasov}, and we will also 
present the solution, for completeness, in the case of D5-brane 
background later on. 

The asymptotic behaviour of $T(R)$ can be found out by using the
properties of hypergeometric function ${}_2 F_1$: 
$\lim_{R \to 0} {}_2 F_1 \to 1$, and $\lim_{R \to \infty} {}_2 F_1 \to 0$. 
Let's analyze the behaviour of the function $T(R)$ and the tachyon potential
$V(T(R))$ asymptotically. 
As $R\rightarrow 0 $, $T(R)\rightarrow  -\infty $: 
\bea
T(R\rightarrow 0)\simeq - R^{p-6}.  
\eea
And, as $R\rightarrow \infty$, $T(R)\rightarrow 0 $: 
\bea
T(R\rightarrow \infty) \simeq R^{{p-7}\over 2}.
\eea
The above can be understood from the simple fact that $p \le 4$,
so $7-p = +ve$.
\footnote{We are excluding the D5 and D6-brane from the above
analysis and for other higher branes the harmonic functions
behave differently}

The effective potential $V(T)$ in various limits behaves as:
\bea
{1\over{\tau_{p'}}}V(T) &\simeq& T^{{(7-p)(p-p'-4)}\over{4(6-p)}},~~~~~~~
R\rightarrow 0 \cr
& \cr
&\simeq& (1+ c T^2)^{{p-p'-4}\over 4}, ~~~~ R \rightarrow \infty.	
\label{eff-pot}
\eea
We would be interested in analyzing the behaviour of the potential
very close to the $Dp$-brane, namely in the region $(R\rightarrow 0)$. 
One can see easily that when $p-p' = 4$, the effective potential 
is a constant. 
When $p-p' < 4, V(T)\rightarrow 0$ so there is a possibility
of the formation of a bound state. And finally, when $p-p' > 4, 
V(T)\rightarrow \infty $, indicating that that the effective potential
is repulsive \cite{burgess2}.

Now, we would like to analyze the rolling of the 
$p'$-brane in  $Dp$-brane background when $p-p' < 4$ and $R\rightarrow 0$.
A classic example would then be the system of $Dp-D(p+2)$ branes. 
We would like to mention that unlike the case of D-NS$5$ system,
where there is no perturbative tachyon between the D-brane and NS5-
brane, in the present case there
is a tachyon in the open string between $Dp$ and $D(p+2)$-branes.  
The bound state is formed due to the dynamics 
of the tachyon in the open string stretching between
the $Dp$ and $D(p+2)$ branes as it essentially stabilises 
the potential \cite{GNS}. In the present case, however, we analyze the
motion of $p$-brane in the background of $(p+2)$-brane where 
the open string tachyon is switched off throughout the dynamical process.
So the 'radial mode' of the brane, which behaves exactly like the 
tachyon and which is the only other field that has been
turned on, would determine the rolling of the probe brane
in this background when both come close to each other. 
Infact, if we start with the system with the perturbative tachyon
turned on and try to analyze the dynamics, it would then become a
coupled system and the above analysis in terms of the radial mode
only is no more valid.

Let's now recalculate some quantities in the Hamiltonian formulation 
from the effective action (\ref{brane}), following the derivation 
of \cite{sen-TT}. The energy momentum tensor $T_{ab}$ 
computed from the action (\ref{brane}) is given by:
\bea
T_{ab} = \tau_{p'} H^{{p-p'-4}\over 4}_p\left[
{{H(R) \p_{a}R \p_{b}R}\over{\sqrt{1+  H(R)\eta^{cd}
\p_{d} R \p_{c} R}}} - \eta_{ab}
\sqrt{1+  H(R)\eta^{cd} \p_{d} R \p_{c} R}\right] 
\eea
The solutions to the equations of motion described by the action
(\ref{brane}) can easily then be computed by working in hamiltonian 
formalism. The momentum conjugate to $R$ is:
\bea
\Pi = {\tau_{p'}{H^{{p-p'}\over 4}_p \p_0 R}\over{\sqrt{1 - H (\p_0 R)^2 
+ H (\vec{\nabla}R)^2}}}
\eea
We can construct the Hamiltonian $H$:
\bea
H &=& \int{d^{p'} x~(\Pi \p_0 R - {\mathcal L})} 
= \int{d^{p'} x ~{\mathcal H}}, \cr
& \cr
{\mathcal H} &=& T_{00} = \sqrt{{{\Pi^2}\over H_p}  + V^2(T)}
\sqrt{1 + H_p {(\vec{\nabla} R)}^2},
\eea
where $V(T) = \tau_{p'}H^{{p-p'-4}\over 4}$. 
The equation of motion for $R$ is given by:
\bea
\p_0 R = {{\delta H}\over{\delta \Pi}} = {\Pi \over{H_p}} 
{{\sqrt{1 + H_p {(\vec{\nabla} R)}^2}}\over{\sqrt{{\Pi^2 \over{H_p}} 
+ V^2(T)}}}.
\eea

As discussed above, we will be interested in the dynamics of the
$Dp'$-brane in the region where 
$T\rightarrow - \infty \> (R\rightarrow 0)$ and $p-p' < 4$. So 
one can ignore the potential $V(T)$ (first line of eqn. \ref{eff-pot})
and the rest of the dynamics is governed by the `tachyon' (that is the
Radial mode $R$). 
The hamiltonian and the equations of motion take the following simpler
form:
\bea
H = \int{d^{p'} x ~{|\Pi|\over{\sqrt{H_p}}}\sqrt{1 + H (\vec{\nabla} R})^2},
\eea
\bea
\p_0 R = {{\delta H}\over{\delta \Pi}} = {1\over{\sqrt{H_p}}} 
{\Pi \over |\Pi|}
\sqrt{1 + H_p {(\vec{\nabla} R)}^2}
\label{dustlike}
\eea
Form eqn (\ref{dustlike}) we get: 
\bea
H_p\left[(\p_0 R)^2 - {(\vec{\nabla} R)}^2\right] = 1
\eea 
When translated in terms of the variables $T$, one gets:
\bea
(\p_0 T)^2 - {(\vec{\nabla} T)}^2 = 1.
\eea
As explained in \cite{sen-TT}, the above equations of motions
when transformed into the old variable $T$, have the natural 
explanations of equations governing the motion of non-rotating
and non-interacting dust. It has been proved in \cite{sen-FT}
that the solutions describing the tachyon on a brane-antibrane
system near the minimum does not admit any plane-wave solutions.
The pressure falls to zero exponentially at very late time as
the tachyon field evolves from any spatially homogeneous
initial configuration towards the minimum.
In the case of D-brane near the NS5-brane, this has also been 
argued in \cite{kutasov}. So it will be interesting to find out 
the field theory of the `tachyon matter' in these set ups.
 
\sect{Homogeneous Solutions}
In this section, we would like to study some homogeneous solutions
to the equations of motion explained in the earlier section. We
assume that the fields in the common transverse directions are
functions of time only $(X^m(t))$. If this
is the case then there is no inhomogeneity in the initial data and
the caustics explained in \cite{felder} don't form at finite time.
Hence the Born-Infeld analysis is appropriate, and the motion of D-branes
can be analyzed by that. Let's assume then the coordinates in the
common transverse directions $X^m$ $(m = p+1,...,9)$, depend only on time.
So the induced metric on the worldvolume of $Dp'$-brane takes the form:
\bea
G_{a b} = H^{-{1\over 2}}_p\eta_{a b} + \delta^0_{a}\delta^0_{b} 
{\dot{X}}^m {\dot{X}}^m H^{1\over 2}_p. 
\eea  
Now substituting the metric and the dilaton into the action (\ref{brane}), 
we get:
\bea
{\mathcal S}_{p'} 
= -\tau_{p'} V\int dt~H^{{p-p'-4}\over 4}_p\sqrt{ 1 - {\dot{X}}^m {\dot{X}}^m
H_p},
\eea
where $V$ is the volume of the $p'$ dimensional space.
It is easy to find out the momentum $P_n$ by varying the above Lagrangian
with respect to ${\dot{X}}^n$:
\bea
P_n = {{\delta {\mathcal L}} \over {\delta{\dot X}^n}} = {{\tau_{p'} V 
H^{{p-p'}\over 4}_p {\dot{X}}_n} \over{\sqrt{1 - {\dot{X}}^m 
{\dot{X}}^m H_p}}}, 
\eea
and the total energy is given by:
\bea
E = P_n {\dot X}^n - {\mathcal L} = {{\tau_{p'} V 
H^{{p-p'}\over 4}_p} \over{H_p\sqrt{1 - {\dot{X}}^m {\dot{X}}^m H_p}}}
\eea
Another quantity of interest is to calculate the Stress tensors 
$T_{\mu \nu}$:
\bea
T_{00} &=& {{\tau_{p'} H^{{p-p'}\over 4}_p} \over{H_p\sqrt{1 - {\dot{X}}^m 
{\dot{X}}^m H_p}}} , \cr
& \cr
T_{i j} &=& -\tau_{p'} \delta_{i j} H^{{p-p'-4}\over 4}\sqrt{1- H_p \dot{X}^m
\dot{X}^m} , \cr
& \cr
T_{0i} &=& 0.
\eea
From the above, it is not difficult to see that the pressure goes
to zero at very late times, so the system behaves like a pressure-less gas.
It would be interesting to work out the case when there is a background
NS-NS flux and/or a worldvolume gauge field, and examine the behaviour
of the pressure $(\propto T_{ij})$ and see how the background flux
is playing a role in the motion of the brane and the eventually the 
time dynamics
of the `tachyon'. Similar issues regarding the D-brane decay in the presence
of a background electric field from the two-dimensional conformal field theory
point of view has been addressed in \cite{sen-DD} and the 
final decay product has been 
shown to be a fundamental string and a ``tachyon matter''. By using the
T-duality transformation on the relevant boundary state, such decay
process has also been studied in \cite{rey-RT,rey-MT}.
It would be nice to repeat the analysis of this paper
in the presence of an electric (magnetic) field and study the time 
evolution process.  

\sect{D-brane dynamics in D5-brane background}

In this section, we would like to analyze the dynamics of D-branes
in the background of $N$ coincident D5-branes where all of them lie
parallel to each other and the worldvolume directions of the probe brane 
falls into the directions parallel to the D5-branes. 
The metric, dilaton and the R-R flux of a system of 
$N$ coincident D5-branes is given by:
\bea
g_{\al \bt} &=& H^{-{1\over 2}}~\eta_{\al \bt}, \>\>\> 
g_{mn} = H^{1\over 2}~\delta_{mn},\>\>\> (\al, \bt = 0,..,5;
\>\> m,n = 6,..,9)\cr
& \cr
e^{2\phi} &=& H^{-1}, \>\>\> F_{mnp} = -\epsilon^r_{mnp} \p_r H. 
\eea
where $H = 1 + {{N g_s l^2_s}\over{r^2}}$ is the Harmonic function
in the four transverse directions of the D5-branes.
The DBI action can be read off from eqn. (\ref{brane}).
The tachyon field and the tachyon effective potential is given by: 
\bea
{dT \over dR} = \sqrt{H (R)} = \sqrt{1 + {{N g_s l^{2}_s}\over R^{2}}},
\label{tachyond5}
\eea
and 
\bea
V(T) = {\tau_{p'} \over{[H(R(T))]^{{p'-1}\over 4}}}.
\eea
In this case however, the T(R) in a monotonically increasing 
function\cite{kutasov}:
\bea
T(R) = \sqrt{Ng_s l^2_s + R^2} + {1\over 2}\sqrt{N g_s} l_s 
\ln {{\sqrt{N g_s l^2_s + R^2} - \sqrt{Ng_s} l_s} \over
{{\sqrt{N g_s l^2_s + R^2} + \sqrt{Ng_s} l_s}}}.
\eea
As $R\rightarrow 0$, $T(R) \rightarrow  -\infty$:
\bea
T(R\rightarrow 0) \simeq \ln R.
\eea
As $R\rightarrow \infty$, $T(R) \rightarrow  \infty$:
\bea
T(R\rightarrow \infty) \simeq R.
\eea
The effective potential in these two asymptotic regions is given by:
\bea
{1\over \tau_{p'}}V(T) &\simeq& 
[\exp T]^{{p'-1}\over 2},\>\>\> T\rightarrow -\infty \cr
& \cr
&\simeq& \left[1 - {(p'-1)\over 4}{{N g_s l^2_s}\over{T^2}}\right],
\>\>\> T\rightarrow \infty.
\eea
One can see that for $p' =1$, the potential goes to a constant in both cases
and the force goes to zero.
In the limit $T \rightarrow - \infty$, which correspond to $R\rightarrow 0$,
one finds that the potential goes to zero exponentially. This is the
expected late time behaviour for the tachyon potential in the rolling
tachyon solutions. One can also compute the energy momentum tensors and
can show that the pressure falls off exponentially at a very late time.
We would like to point out that the above analysis
is valid for the region $Ng_s \gg 1$ where the description in terms of
closed string background makes sense. The systematics of the D-brane
motion in this D5-brane background can be analyzed by following 
\cite{kutasov}.  
In this case, it is much simpler to show the absence of plane wave
solutions near the minimum of the potential $V(T)$.
It is also somewhat straightforward to show the possible trajectories 
of the probe brane in the directions transverse to the D5-brane, after
passing to the polar coordinates, following the derivations of \cite{kutasov}.
We however skip those details here. 

\sect{Conclusion}

In this paper, we have studied the D-brane dynamics in the background
that is generated by a stack of static branes from the point of view of
the DBI action. The study of real time dynamics of the D-brane
provides an example of time dependent process in string theory. We have    
shown that when there is no WZ interaction between the probe and the
background branes, one can map this problem into a similar problem of
tachyon condensation and study the time evolution process. We discuss
the classical Hamiltonian formulation and study some homogeneous solutions
to the equations of motion. We study the interaction potential and show
that the pressure goes to zero at very late times. Finally we remark on
D-brane dynamics in the presence of D5-brane
which shows that the pressure goes to zero at late times exponentially.
There are further questions which needs to be addressed. Though
the analysis of brane dynamics can be mapped, in these simple cases,
to a problem of real time tachyon
dynamics of the open string theory, it is not clear how all the effects
like the gauge fields and the other R-R fields could be incorporated
in these studies. For example: it has been discussed in \cite{burgess2}, 
that a possible WZ interaction could arise between the background 
and the probe brane
when the dimensionality between the two differs by six: $( p+p' = 6)$.
The D-brane trajectory has been shown, in that context, to be 
equivalent to that of a
non-relativistic dyon in a magnetic monopole background. So it would
be interesting to repeat the above analysis to notice the
changes in the energy momentum tensors and study the time evolution
process. One way, however, is to construct boundary state
of the systems and study the properties of stress tensors to
analyze the behaviour of `tachyon matter'.
Another interesting aspect would be to discuss 
the D-brane dynamics in more detail in the presence of various
other non-perturbative objects and learn more about the time
dependent classical solutions of string theory.
Of course an interesting exercise would be lift the
discussion of the DBI to that of the full conformal field theory analysis.
In that picture, the properties of the classical time dependent
solutions would correspond to perturbing the original D-branes 
by an exactly marginal deformation and study its effects.
More recently the boundary states of
rolling D-branes in NS5-brane background are constructed in \cite{BS-RD}. 
So it will be interesting to construct, if possible, more generalized 
boundary states of Rolling D-branes in the presence of various flux. 
In the brane picture it would be 
useful to show how these problems can be analyzed from more generalized 
set ups and the possible connections with the cosmology (a related 
problem has been addressed recently in \cite{yavar}). 
We hope to return to some of these issues in near future.       

\vskip .2in
\noindent    
{\large\bf{Acknowledgment:}} 
I am grateful to A. Sen for suggestions on the manuscript and for
various useful discussions. I would like to thank M. Bianchi, 
J. David, J. F. Morales, R. R. Nayak and especially A. Sagnotti 
for numerous interesting discussions. This work was supported in 
part by I.N.F.N., by the E.C. RTN programs HPRN-CT-2000-00122 
and HPRN-CT-2000-00148, by the INTAS contract 99-1-590, by the MURST-COFIN 
contract 2001-025492 and by the NATO contract PST.CLG.978785.

\end{document}